# Hallucination vs interpretation: rethinking accuracy and precision in AI-assisted data extraction for knowledge synthesis

Xi Long[1], Christy Boscardin[2], Lauren A. Maggio[3], Joseph A. Costello[4], Ralph Gonzales[4,5], Rasmyah Hammoudeh[5], Ki Lai[6], Yoon Soo Park[3], Brian C. Gin[3,7]*

Author Affiliations:

1. University of Illinois Chicago, Department of Biochemistry and Molecular Genetics
2. University of California, San Francisco, Department of Medicine and Department of Anesthesia and Perioperative Care
3. University of Illinois Chicago, Department of Medical Education
4. University of California, San Francisco, Department of Internal Medicine
5. University of California, San Francisco, Clinical Innovation Center
6. University of California, San Francisco, Health Information Technology
7. University of Illinois Chicago, Department of Pediatrics

*Correspondence: Brian C. Gin, email: bgin@uic.edu

## Abstract

Introduction: Knowledge syntheses (a.k.a. literature reviews) are integral to health professions education (HPE), consolidating findings across multiple publications to advance theory and practice. However, they are highly labor-intensive, especially during the data extraction phase. Artificial Intelligence (AI)-assisted extraction methods promise efficiency yet raise concerns about accuracy, making it crucial to distinguish AI-generated "hallucination" – incorrect or fabricated content – from legitimate interpretive differences resulting from subjective judgment.

Methods: We developed an extraction platform utilizing large language models (LLMs), to automate data extraction. Extraction accuracy and error rates were evaluated by comparing AI to human extraction responses across 187 publications and 17 extraction questions from a previously published scoping review. We measured AI-human, human-human, and AI-AI consistencies using interrater reliability (for categorical responses) and thematic similarity ratings (for open-ended responses). Errors were identified by comparing extracted responses to source publications and qualitatively characterized.

Results: AI extraction was highly consistent with human responses for concrete questions explicitly stated in manuscripts (e.g., title, aims), and was lower for questions requiring subjective interpretation or not explicitly addressed (e.g., Kirkpatrick's outcomes, study rationale). Remarkably, human-human consistency was not higher than AI-human consistency and reflected the same pattern of question-dependent variability. Analysis of discordant AI and human responses (769/3179 = 24.2%) revealed that AI inaccuracies were responsible for only a small number of cases (1.51%), while interpretive differences – multiple correct answers, unequal levels of detail, and minor classification disagreements – were responsible for the bulk of AI-human discord (18.3%). Humans (4.37%) were nearly 3 times more likely to state inaccuracies than AI.

Discussion/Conclusion: Our findings suggest that AI's accuracy in data extraction is predominantly influenced by interpretability rather than hallucination. Repeating AI extraction can help identify extraction questions subject to (desired) interpretative complexity or (undesired) ambiguity, enabling refinement before human involvement. AI demonstrates potential as a transparent, trustworthy partner in knowledge synthesis, though researchers must remain cautious about over-reliance on AI interpretations, which may neglect critical human contextual insights, expertise, and positionality.

## Introduction

Knowledge syntheses play a key role in health professions education (HPE) scholarship (Maggio et al., 2020). Also known as literature reviews, knowledge synthesis consolidates findings across publications, defines terms, contrasts theories, and helps develop new theories. In HPE, researchers conduct a variety of knowledge synthesis types (e.g., scoping reviews, systematic reviews, critical reviews, realist reviews) (Maggio et al., 2020). While protocols between these knowledge synthesis types differ, data extraction (or data collection) – the process of systematically gathering information from each included publication – plays a key role in all of them.

Data extraction is one of the most labor-intensive steps in a knowledge synthesis (Pham et al., 2018), requiring human team members to answer extraction questions (also called elements or fields) about each manuscript multiple times. Advances in artificial intelligence (AI) have led to the development of AI extraction platforms or strategies that have promised to alleviate some of this burden (Bolaños et al., 2024). At the same time, concerns about accuracy (hallucinations), bias, and unclear reflexivity/positionality temper expectations for how AI can contribute to the research process (Gin et al., 2025). Further, AI overreliance may neglect humans' contextual insights or content expertise. This raises the questions of how an HPE researcher determines whether to use an AI extraction platform in their research, and if so, which steps AI may be allowed to contribute to, and how deeply.[1]

Recent AI extraction strategies rely on large language models (LLMs) to process publications and automatically provide answers to extraction questions from them (Bolaños et al., 2024). Spillias, et al. compared a commercial extraction platform to an LLM-based prompting strategy to determine whether AI could reliably assess whether publications contained answers to specific extraction questions, finding only modest agreement with human extractors (Spillias et al., 2024). Their and others' findings (Bernard et al., 2025) illustrate the central concern of any AI-based tool, that it may "hallucinate" – in the case of extraction, this would mean generating factually inaccurate extraction data that cannot be grounded in the source publications.

However, the distinction between a "hallucination" and interpretation is not always clear. Even if an AI were to be completely accurate[2] in extraction – that is, all the data it extracted could be verified against the source publications – there still could be extraction questions that may be subject to interpretation. For example, interpretable questions may have more than one justifiable answer, invoke a subjective rating, or involve a judgement call. While superficially, extraction may appear as an objective practice not subject to researcher interpretation, it is influenced by the same reflexivity and positionality considerations that shape how researchers approach and interpret qualitative analysis.

The expectation and tolerance for such interpretative variability (hereafter, "interpretability") may also depend on the review type. For example, while systematic reviews conducted from a positivist (or post-positivist) paradigm may involve extraction questions with less inherent interpretability, narrative or critical reviews conducted from a subjectivist paradigm may involve extraction questions with greater interpretability. Regardless of review type or intended philosophical stance, an AI extraction platform would need to be transparent about the degree to which its response represents an interpretation, describe how it used the source to make that interpretation, and avoid biases that may skew such interpretations.

We can frame concerns about AI accuracy and bias in the concept of AI trustworthiness (Ammanath, 2022; Baker-Brunnbauer, 2023; U.S. Department of Health & Human Services, 2021). Trustworthiness involves ability, integrity, and benevolence (Gin et al., 2024; Mayer & Mayer, 2024). With respect to *ability*, an AI platform for extraction would need to generate responses consistent with human responses to the same extraction questions. *Integrity* relates to the transparency of how an AI extraction platform generates its results. A transparent AI extraction tool would provide evidence for its results (including quotes from sources, estimates of certainty, and associated reasoning steps), and enable access to the underlying AI models and generation schema (Grimmelikhuijsen, 2023). Finally, a *benevolent* AI extraction tool would adhere to the ethical principles and motivations of the researchers, reflecting their contextual priorities, positionalities, and minimizing biases. As such, an AI platform would need to be able to either reflect the researchers' viewpoints, a different viewpoint, or to clearly delineate its own viewpoint (Cook et al., 2025).[3]

---

[1] Indeed, recent tools claim to completely automate the process of performing research (i.e. Perplexity AI, Google AI, Elicit, and OpenAI all offer some type of "deep research" tool).
[2] Acknowledging that the post-positivist notion of "accuracy" does not conceptionally align with the subjectivist notions of positionality and reflexivity.

The overarching aim of this study is to provide evidence for researchers to make informed decisions about the potential use and interpretation of AI-generated results during the data extraction phase of a knowledge synthesis. Thus, **the goals of this study are to assess the consistency between AI- and human-based extraction, and to determine whether discordance reflects AI hallucinations or alternative interpretations of extraction questions.** To set a standard for comparison, we measure the consistency of responses provided by different human extractors to the same question/publication pairings. To address hallucination vs interpretability, we qualitatively and quantitatively characterize discordant AI and human responses, referencing the original publications as the source of truth. We perform these analyses as a case study to reproduce an extraction table from a published scoping review – for which we developed an AI-based extraction platform to maximize AI trustworthiness. Our findings can provide researchers with a foundation to contextualize and interpret the results of AI-based extraction, to assess both its potential value and risk to their research and creative processes.

## Methods

*Overview.* To address our study aims, we developed a platform utilizing LLMs to perform data extraction from publication texts, and assessed its consistency relative to human-performed extraction. We used the platform to reproduce an extraction table from a previously published KS described below. To measure the consistency between AI and human extraction (*AI-human consistency*), we compared their interrater reliability (for extraction questions with categorical responses) or thematic similarity (for extraction questions with free text responses). To define a standard of consistency, we assessed *human-human consistency* by performing a targeted, manual re-extraction on a subset of the extraction table, measuring the consistency of human responses on the same publication-question pairings. To assess the variability of AI generated responses (*AI-AI consistency*), we compared responses from repeated AI-performed extractions for the same question/publication pairs. Altogether, we assessed three measures of consistency: AI-human consistency, human-human consistency, and AI-AI consistency. To assess for AI hallucinations (i.e. factual inaccuracies) we performed a detailed analysis of discordant AI and human responses to identify and characterize errors and inaccuracies relative to the ground truth of each publication's text.

---

[3] While some may argue that AI models do not have positionality, we do know they are subject to reproducing biases – and that reflexivity and positionality are related to both implicit and explicit biases.

*Data sources and extraction questions*. The extraction table (questions and extracted data) selected for reproduction was selected from a scoping review on AI in medical education (Gordon et al., 2024). This openly published extraction table consisted of N=187 full-text manuscripts and 17 extraction questions (8 categorical, 9 free response), representing a total of 3179 individual responses. This KS was selected for the case study because it included both straightforward, as well as more interpretable extraction questions. Further, the subject of this KS did not require highly specialized context-specific knowledge that would have required us to provide extensive explanations of terms, settings, or processes to the AI.

The extraction prompts we developed – used to provide instructions for the LLM(s) to answer each extraction question – are shown in Table 1 (the full prompts, including the possible responses for categorical questions are provided in Appendix 1). We categorized these questions into categorical and free response, denoted in C or F, respectively.

*Extraction platform*. We developed MAKMAO (Machine-Assisted Knowledge extraction, Multiple-Agent Oversight), a platform (Appendix 2) to extract data from documents using AI large language models (LLMs). While commercial products (also based on LLMs) for extraction are also available, our aims require us to establish control over how (and which) LLMs are used to extract data, which is not currently possible with proprietary platforms. This level of control and transparency is necessary for us to gauge our entrustment of the AI tool, and to make meaningful comparisons of different LLMs and response pooling strategies.

**Table 1:** Extraction questions. "…" indicates that the prompt was abbreviated to fit in this table (see Appendix 1 for full-length prompts).

| Extraction question abbreviation | Extraction question prompt excerpt (see Appendix 1 for full prompt) | Categorical [C] or **Free response** [F] |
|---|---|---|
| Title | What is the title of the paper? | F |
| Country of origin | In what country(ies) was the study performed? | F |
| Aims | What are the aims, goals and objectives of the paper? | F |
| Results | Summarize the results of the study ... | F |
| Implications | State what the authors proposed as the study's implications ... | F |
| Specialty | What medical specialty does the paper apply to (if applicable)? ... | F |
| Study design | Identify the study design and categorize it into a concise label based on the methodology used. ... | F |
| AI rationale | What was the study's rationale for using AI? ... | F |
| NLP | Did the study utilize natural language processing (NLP)? ... | C |
| Language model | If a paper used an AI language model or LLM please provide the name(s) of such models. ... | C |
| AI application | What area of medical education did the study apply AI to? ... | C |
| AI method | What AI method was used? ... | C |
| AI use case | What analytic or technological purpose did the study use AI to perform? ... | C |
| Stage of education | What was the stage of education? ... | C |
| SAMR | Use the SAMR framework[a] to describe how the study applied AI to the medical education task. | C |
| Kirkpatrick's outcomes | If the study explicitly measures learning outcomes, determine which Kirkpatrick evaluation level the humans' outcomes are most consistent with. ... | C |

[a] SAMR Framework – Substitution (using AI as a direct substitute for traditional methods without functional change), Augmentation (using AI as a substitute with functional improvements), Modification (using AI enables significant task redesign), or Redefinition (using AI allows for the creation of an entirely new, previously inconceivable task)

*Data extraction with LLMs.* For every publication, MAKMAO answered each extraction question by systematically applying a prompt derived from the extraction question (Table 1) to the full text of the publication (Figure 1a). We optimized the text of each extraction prompt iteratively until we could no longer achieve further improvement in the AI-human consistency. Optimizations included refining the question, providing definitions of terms, and explaining responses options (for categorical questions).

We utilized Python 3.12 to automate a structured prompting strategy that required the LLM to perform a series of steps for each extraction question (Figure S1):

1) Search the publication text for quotes relevant to the question
2) Provide a detailed rationalization for how the question is to be answered from the quotes (including if the answer cannot be found in the publication text)
3) Provide an answer to the question based on the quotes

This structured prompting strategy was enforced by dynamically constructing a nested Pydantic class specific to each set of extraction questions. This results in a structured JSON output format for each set of questions, with the output for each question containing generated responses to the steps above (Table 2a). We tested this scheme with closed parameter LLMs (GPT-4o and GPT-4o-mini), a closed parameter large "reasoning" model (LRM[4]) (GPT-5), and an open parameter LLM (qwen2.5:14b).

**Figure S1**. The chain-of-thought prompting strategy used for an LLM to answer each question about a given publication.

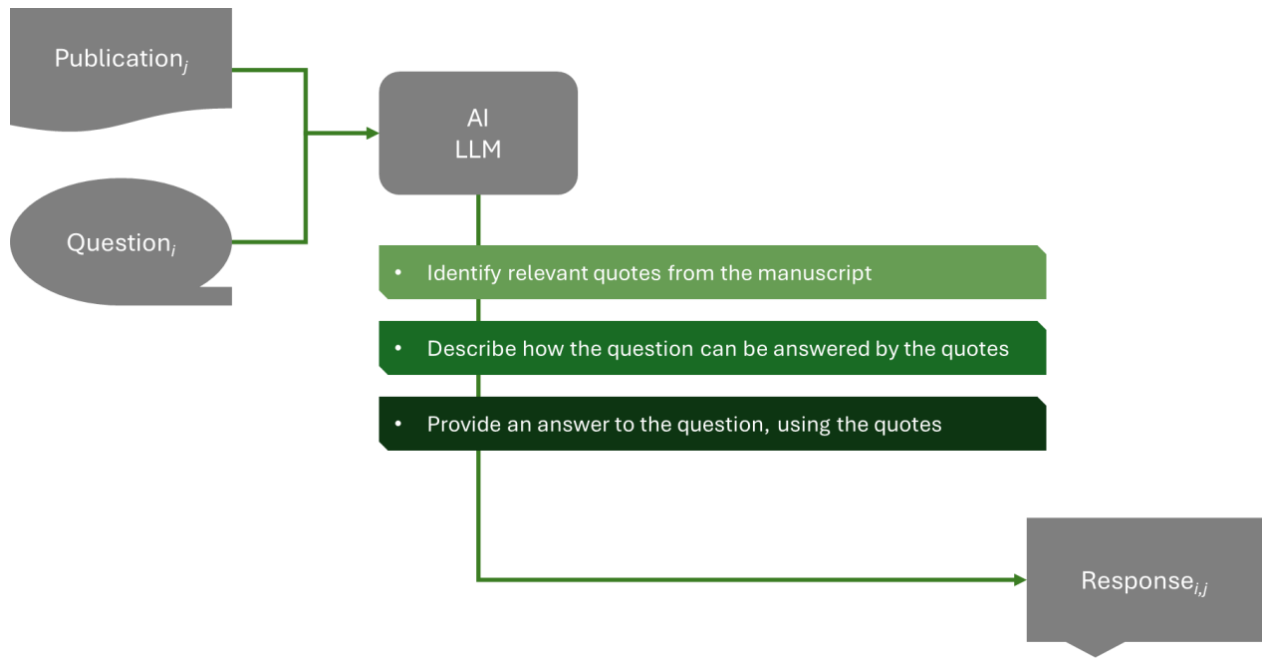

*Reflexivity framing question responses*. MAKMAO provided a mechanism for users to specify system-level instructions to the underlying LLM – i.e. a special prompt that sets the LLM's apparent behavior when answering prompts. This allowed the user to provide the LLM with definitions and descriptions of theoretical frameworks, epistemologies, and ontologies that are necessary to lay out a pragmatic overview of a research paradigm – helping to shape the responses the LLM will generate. The context we provided for this study framed it in the medical education literature but did impose a specific epistemology or theoretical stance. This was deliberately chosen to assess baseline LLM responses. Specific definitions pertaining to each question (e.g. Kirkpatrick levels) were provided in the extraction question prompts.

---

[4] Large reasoning models (LRMs) are a type of large language model (LLM) that are trained to generate and use an explicit, step-by-step reasoning process when responding to prompts (Xu et al., 2025). This approach is related to Chain-of-Thought (CoT) prompting, but differs in that LRMs are trained with objectives that also evaluate and optimize the quality of their reasoning steps during training, not just the final answer.

*Measuring response consistency*. To compare the consistency of AI extraction to human extraction, we calculated the human-AI interrater reliability (for categorical questions, using Cohen's kappa) (Park et al., 2016) or "thematic similarity" rating (for free response questions) of each extraction question over the entire set of data. Thematic similarity was first assessed by asking human raters, BG and XL, to compare AI and human-extracted responses, to develop a simple "thematic similarity rating instrument." After scoring the thematic similarity between AI and human responses of half of the free-response data manually in this manner, we automated this scoring using AI, finding it to be consistent with human rating (See Box 1 and Appendix 2). While interrater reliability and thematic similarity ratings are related but nonequivalent measures, they were similar with respect to both scale and range. Thus, we compared these two different measures of AI-human consistency to evaluate categorical and free response questions on the same numerical scale (from 0 to 1, where 0 represents no consistency, and 1 represents complete consistency).

*Human extraction*. To assess human-human consistency (and define a standard to compare AI-human consistency), we needed two sets of human-extracted responses for each question/publication pairing. Thus, we performed a targeted, manual extraction of 5 (categorical and free response) questions (representing the low, moderate, and high AI-human consistency groups as described in the Results) over a subset of 30 manuscripts (each result was extracted and confirmed by two authors XL and BG).

*Response variability and uncertainty*. Given their probabilistic nature, LLMs may generate different answers to identical prompts (i.e. pairings of extraction questions and publications) when starting from different random seeds (i.e. numerical values used to initialize an LLM's random starting point – since digital random number generators are inherently deterministic) (Gin et al., 2025). In practical terms, an LLM may return a different response each time it is prompted to answer a given question/publication pairing. We hypothesized that this *response variability*,[5] relates to the LLM's uncertainty in the answer to that question – and may stem from question interpretability and/or hallucinations.

*Leveraging AI response variability to flag interpretability and hallucination*. We designed an additional multi-agent scheme (Figure 1b) to alert the user to response variability of any particular question/publication pairing, which can be indicative of question interpretability and/or hallucination. To elicit response variability, set model temperature to 0.15 and utilized GPT-4o-mini, a lightweight LLM that allows multiple independent sessions to be run simultaneously.[6] In this scheme, multiple independent LLMs running in tandem answer each question/publication pairing, and their responses are compared to each other and to the original publication text by an additional LLM (Figure S2). This process can lead to 1) a "consensus" response for a given question/publication pairing (or reveal a lack of such consensus), and 2) detection of hallucinations if a given response is not grounded in the source text (Table 2b).

---

[5] Generating different responses to the same prompt, but starting from different random seeds while holding other parameters constant, including temperature, top-K, etc.

*Discordance analysis and hallucination detection*. To further unpack the sources of discordance between AI and human responses, we identified errors by independently comparing discordant AI and human responses to the source publications, without comparing them to each other. Treating each publication as the source of truth, we used an LLM (GPT-5-mini) external to our platform to independently compare the AI (GPT-4o) and human responses to the publication text for each extraction question. We examined each response for statements that were counterfactual to the source publication, and for classifications or inferences that could not be justified by the source publication. We organized our findings by extraction question type of error exhibited by the AI or human response, developing a classification scheme by performing (BG and XL) a qualitative content analysis (Hsieh & Shannon, 2005) on the types of extraction errors exhibited (Table 4). Our initial codes focused on a broad classification of error into *inaccurate* and *insufficient* extraction, which we then refined into sub-codes. The sub-codes included:

- **O**mission = failure to identify a major element being extracted
- Major mis**C**lassification = true misclassification, improper equating
- **F**actual lapse = assertion that the publication contains facts that it does not state
- **S**pecificity deficit = insufficient detail, lack of completeness, oversimplification, or generalization
- **M**inor misclassification = classification into a level neighboring (or similar to) the most appropriate level

---

[6] Note that this strategy is less useful with LRMs (such as GPT-5) because they do not allow the model temperature to be tuned and are generally more deterministic in their responses.

**Figure S2**: Screening for hallucinations via an additional LLM that compares a given response to the original publication text.

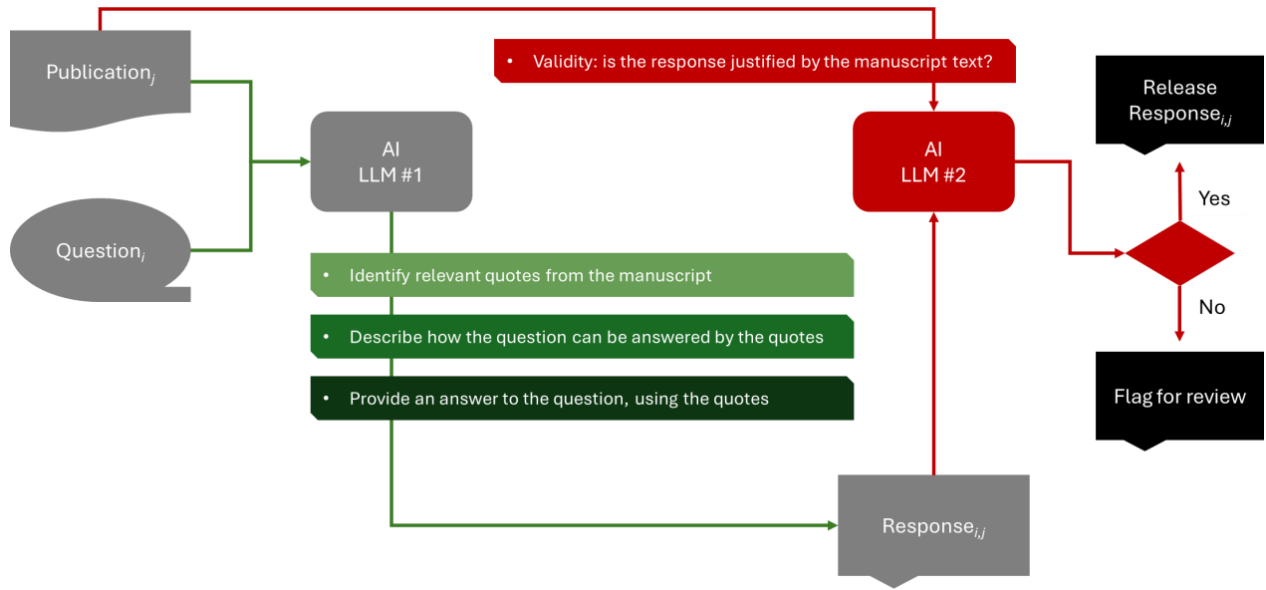

**Figure 1:** Flowcharts showing LLM-based extraction schema. For a single (a) extractor, each question/publication pair (*i,j*) is presented to an LLM, which is prompted to answer the question in the format specified in Table 2a, and the results aggregated in a table (with columns representing each question *i*, and rows representing each publication *j*). For multiple (b) extractors, the process is repeated with several independently acting AI LLMs (i.e. "agents") each performing the extraction as in Figure 1a. Their results are then aggregated and compared by another independent LLM (LLM$_C$), which produces the responses specified in Table 2b. This results in an estimate of certainty for each question/publication pair, based on the similarity of the replicated extraction responses.

(a) Single "extractor" scheme:

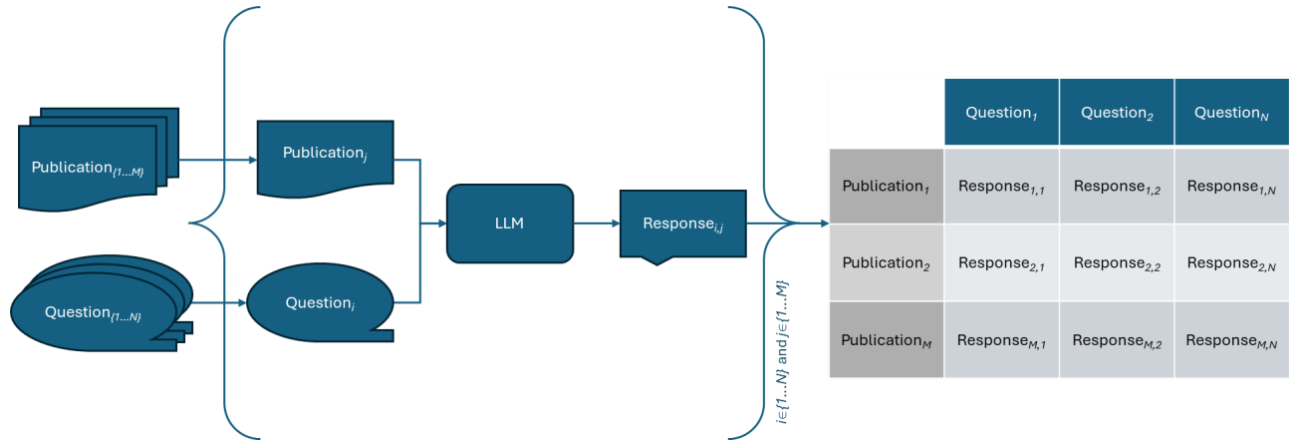

(b) Multiple "extractor" scheme:

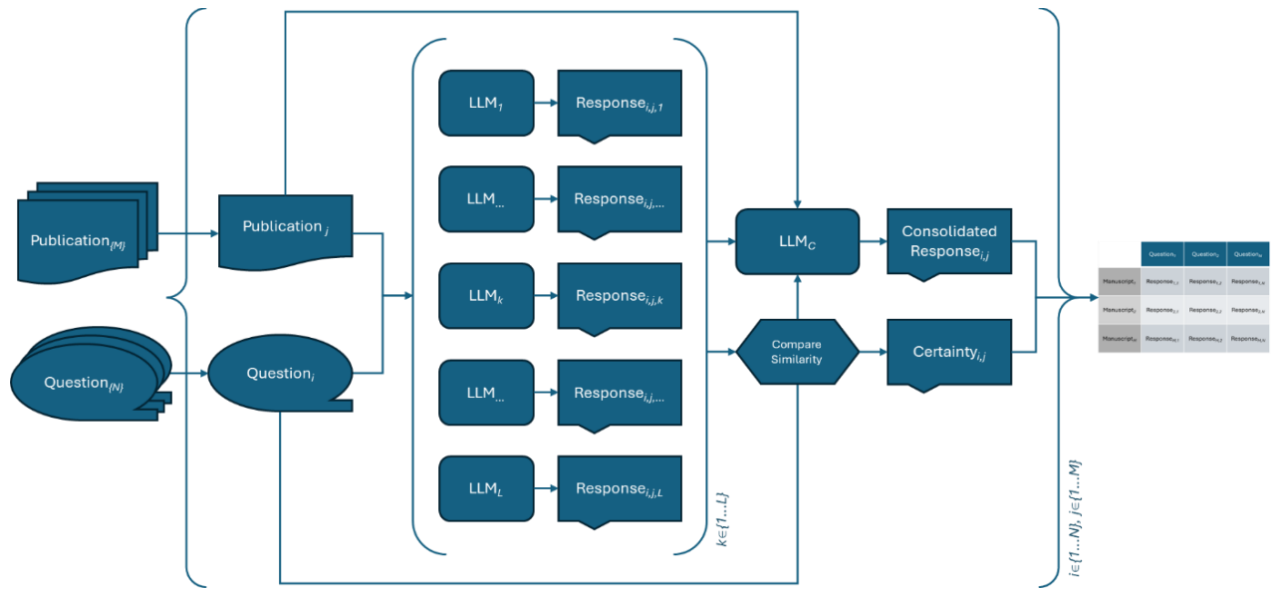

Table 2: Extraction data generated for each Question + Manuscript pairing (Response$_{i,j}$).

| Single "Extractor" Scheme (a) | Multiple "Extractor" Scheme (b) |
|---|---|
| Short answer | Short answer |
| Long answer explanation | Consolidation explanation + quotes |
| Supporting quotes | Level of certainty |
|  | Individual responses from each extractor |

**Box 1:** Thematic similarity rating instrument. To compare free response questions for their thematic similarity, we developed the following simple scoring scheme, which we initially applied with human raters (over all 187 manuscripts), then automated by AI as we refined our extraction questions.

> Compare the following two responses and score them on a scale of 0 to 1, where:
>   0 = thematically dissimilar
>   0.5 = thematically similar with differences in details
>   1 = thematically similar with similar details
> Response 1: …
> Response 2: …

## Results

### AI vs Human Performance - evidence for question-specific interpretability

*AI-human consistency.* We first assessed AI's ability to reproduce human responses by comparing its extraction responses to answers provided by humans over all 17 extraction questions and 187 publications (Figure 2). We found that AI produced answers highly consistent with humans' responses for questions related to publications' titles, aims, and results. However, AI exhibited less consistency with human responses for questions requiring characterization of a study's Kirkpatrick's outcomes, AI rationale, and overall study design.

AI-human consistency (defined as the interrater reliability or thematic similarity rating for categorical and free-response questions, respectively) varied similarly for responses to both categorical (n=8) and free-response (n=9) extraction questions, showing a range of 0.25 to 0.99 for interrater reliability and 0.45 to 1.0 for thematic similarity rating. We categorized extraction questions by their tendencies to elicit responses within a specific range of AI-human consistencies in Table 3 – high, moderate, and low. In general, differences between the AI-human consistency (as assessed by these measures) appeared to be statistically significant when comparing extraction questions across each of these groups (i.e. the 95% confidence intervals did not overlap between groups).

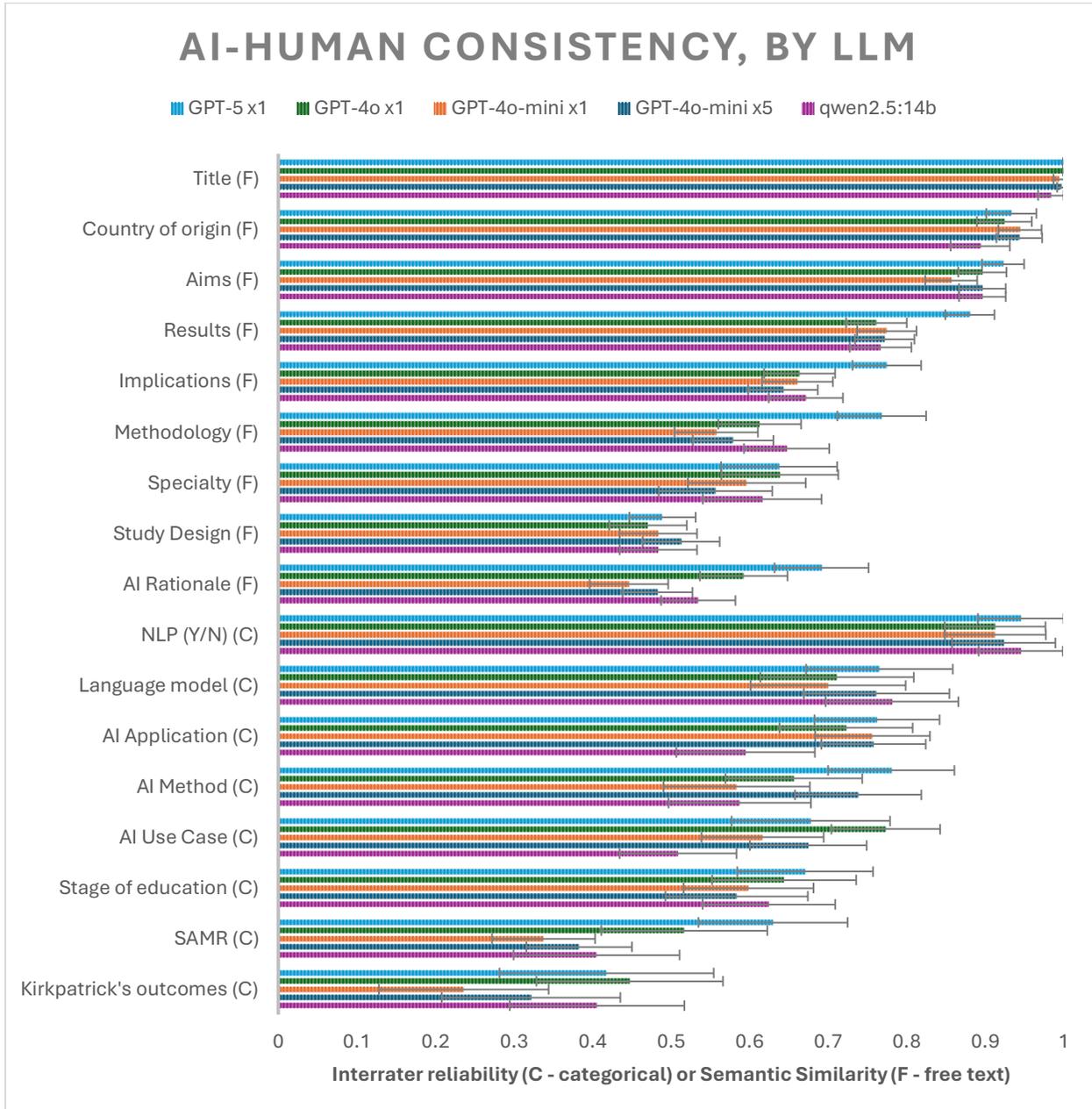

**Figure 2:** *AI-Human consistency demonstrates a strong question-dependent pattern across different LLMs*. The bar graph shows the consistency between human and AI responses over all manuscripts, grouped by extraction questions. For categorical questions, the consistency is measured by interrater reliability (Cohen's Kappa) and for free response questions by a thematic similarity rating (see Methods). The 95% confidence intervals for each consistency measure are shown by the brackets. The bottom half shows results for categorical questions (C), while the top half for free response questions (F). Response variability is observed

across both question types, with questions requiring interpretation demonstrating lower AI-human consistency. We applied different AI models to perform the extraction (colored bars): GPT5, GPT-4o, GPT-4o-mini, GPT-4o-mini run 5 times, and qwen2.5:14b. Each LLM demonstrated a similar question-dependent pattern of consistency with human responses (Pearson correlations between them ranged from 0.91 to 0.98).

Table 3: Categorizing extraction questions by AI-human consistency

| High AI-human consistency (interrater reliability or thematic similarity rating >0.75) | Moderate AI-human consistency (0.5-0.75) | Low AI-human consistency (<0.5) |
|---|---|---|
| • AI application [C]<br>• Results [F]<br>• Aims [F]<br>• NLP (Y/N) [C]<br>• Country of origin [F]<br>• Title [F] | • Specialty [F]<br>• AI use case [C]<br>• AI method [C]<br>• Methodology [F]<br>• Stage of education [C]<br>• Implications [F]<br>• Language model [C] | • Kirkpatrick's outcomes [C]<br>• SAMR rating [C]<br>• Study design [F]<br>• AI rationale [F] |

Questions eliciting responses with AI-human consistency appeared to require little interpretation (such as title and country of origin) or reflect interpretations that authors had written explicitly in the publications (such as aims, results, and AI application). Questions eliciting lower consistency appeared to require interpretation of the study (SAMR, Kirkpatrick's outcomes), or pertain to topics stated less explicitly in the publications (study design, AI rationale).

*Comparing AI-human consistency across AI models.* Using MAKMAO, we compared the AI-human consistency of different AI models (GPT-5, GPT-4o, GPT-4o-mini, and qwen2.5:14b) for these extraction questions (Figure 2, bar colors). The models represented both closed and open-parameter LLMs of various sizes (from 7 billion to 70+ billion parameters), and included one "reasoning" (LRM) model (GPT-5). Additionally, we compared whether running a "small" LLM[7] (GPT-4o-mini) multiple times in parallel (Figure 1b), followed by consolidation of these responses, would improve AI-human consistency.

Our results showed that the AI models exhibited a similar pattern of question-dependent AI-human consistency reflecting the classification in Table 2. While the LRM (GPT-5) demonstrated higher AI-human consistency than all LLMs on the results, implications, AI rationale and SAMR questions ($p < 0.05$),[8] there were no statistically significant differences

---

[7] Generally, an LLM with less than 70 billion parameters. GPT-4o-mini is thought to have 8 billion parameters.
[8] The higher performance of the LRM, GPT-5, is likely artefactual and not related to potential superiority of LRMs to LLMs. The apparent higher performance of GPT-5 appears to be attributable GPT-5 having ingested

between LLMs themselves (GPT-4o, GPT-4o-mini, GPT-4o-mini x5, and qwen2.5:14b).[9] All of these models demonstrated a relatively high correlation between their AI-human consistencies, suggesting that they all struggled to answer certain questions consistently relative to the human standard (Pearson's correlation ranged from 0.91 to 0.98, considering the "4o-mini x5" model as baseline).

*Human-human consistency.* We considered whether the interpretability of an extraction question may lead to a "ceiling effect" on the consistency of repeated extraction, whether by AI or humans. Since the question-dependent AI-human consistency did not significantly improve with more capable AI models or prompting strategies, we hypothesized that the upper bounds of AI-human consistency may not reflect a limitation of AI's ability replicate human extraction, but rather, that less-consistently answered questions may be more *interpretable* (e.g. that it may have more than one justifiable answer, or a range of plausible answers). Thus, we asked whether responses from multiple human extractors (i.e. human-human consistency) would also reflect this question-specific pattern of response consistency. To provide two sets of human extractions for the same questions, we manually performed a targeted re-extraction as described in the *Methods*.

---

this KS (and its data) during its training, which included publications through June 2024 and did include that study (See Appendix 3). As such, we excluded it from further analysis.

[9] These comparisons were all performed with optimized prompts (see supplemental) that had already gone through iterative refinement to improve AI-human consistency (as described in Methods, such as including definitions of terms/concepts and providing optimized categories) based on the original questions in the extraction table. In general, prompt refinement appeared to improve the responses generated by smaller LLMs more than larger, more capable LLMs—which appeared to be less sensitive to prompting strategy, likely because they contain more foundational knowledge and "reasoning" capability (data not shown).

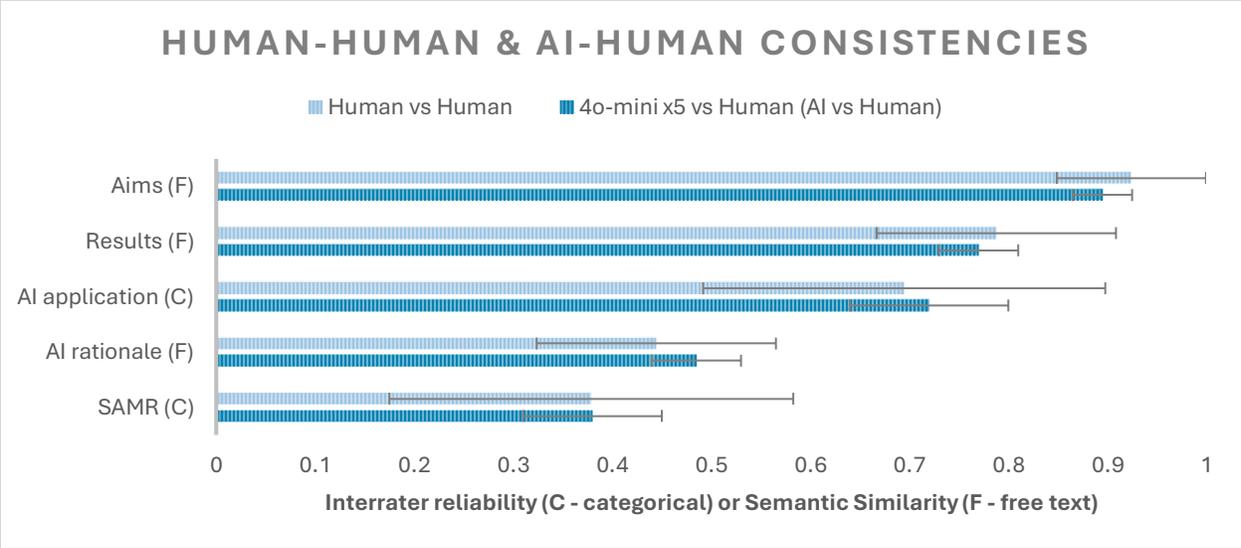

**Figure 3:** *Human-human consistency is indiscernible from AI-human consistency, demonstrating the same question-dependent pattern.* The bar graph shows the consistency (IRR for categorical C, and semantic similarity for free text F) between two sets of human-performed extraction responses (Human vs Human) to the same question/publication pairings over a targeted subset of the data shown in Figure 2. This is compared to AI-human consistency over the same subset (4o-mini x5 vs Human). The brackets show the 95% confidence intervals for the consistency measures. Humans were no more consistent with each other than AI was consistent with humans. There is no statistically discernable difference between the consistency of human vs human and AI vs human extraction, and they demonstrate a similar question-specific pattern (Pearson correlation 0.99 ± 0.08).

We found that the human-human consistency (as similarly measured by interrater reliabilities and thematic similarity ratings) for this subset of extraction questions replicated the question-specific pattern we had seen with the AI-human consistency (Figure 3). The two human extractor groups exhibited low consistency answering the same questions that AI had also answered less consistently relative to the human responses (AI rationale, SAMR), while the humans exhibited high consistency answering questions for which the AI had consistently replicated human responses (aims, results). In other words, ***humans also had difficulty consistently answering the same questions that AI struggled to answer consistently.***[10]

Statistically framed, within extraction questions there were no significant differences in the responses' human-human consistencies versus human-AI consistencies, while between extraction questions there were significant differences. This question-specific pattern was mirrored in the human-human and human-AI responses (Pearson's correlation comparing human-human consistency with human-AI consistency was 0.99 ± 0.08). These findings indicated that question-specific response variability was present irrespective of whether the extractor was human or AI – suggesting an inherent interpretability associated with each extraction question.

---

[10] Interestingly, the AI-human consistency of the LRM (GPT-5) was significantly ($p<0.05$) higher than the human-human consistency for the AI rationale and SAMR questions, while still demonstrating high correlation to the human-human pattern (0.97). This is likely an artefact caused by GPT-5's prior knowledge and training on Gordon, et al.'s KS. See Appendix 3.

*AI-AI consistency.* We considered whether question interpretability could also be reflected in the range of possible AI responses to each question. To examine whether AI alone (without comparison to human responses) could provide insights on how interpretable an extraction question may be, we leveraged LLMs' generative variability. A single LLM may generate different responses to the same prompt each time it is instanced,[11] owing to several factors, including that its training data invariably contains sources representing different viewpoints. Thus, we hypothesized that if the question-specific patterns of AI-human and human-human consistency reflected question interpretability, this would manifest similarly in the AI-AI consistency, with questions of higher interpretability leading to lower AI-AI consistency and vice versa.

We assessed AI-AI consistency by using the same AI model (4o-mini x5) to answer the same extraction questions on the same manuscripts twice (Figure 4). Somewhat expectedly, we found that the AI-AI consistency was higher (on average) than both the AI-human and human-human consistencies.[12] The pattern of question-dependent consistency (seen in the AI-human and human-human consistencies) remained: the Pearson correlation between the AI-AI and AI-human consistencies over all 17 extraction questions was 0.85 ± 0.14 (Figure 5). This suggested that by using AI models alone (with repeated extractions), one could identify extraction questions with greater interpretability based on their lower AI-AI consistency (and vice versa), without the need to compare to human responses.

---

[11] i.e. each time an LLM is reloaded into memory without a prior message/chat history, but starting with a different random seed but keeping other parameters (such as temperature and top-k) equal.
[12] Although, the AI-AI consistency can be decreased by increasing the model temperature.

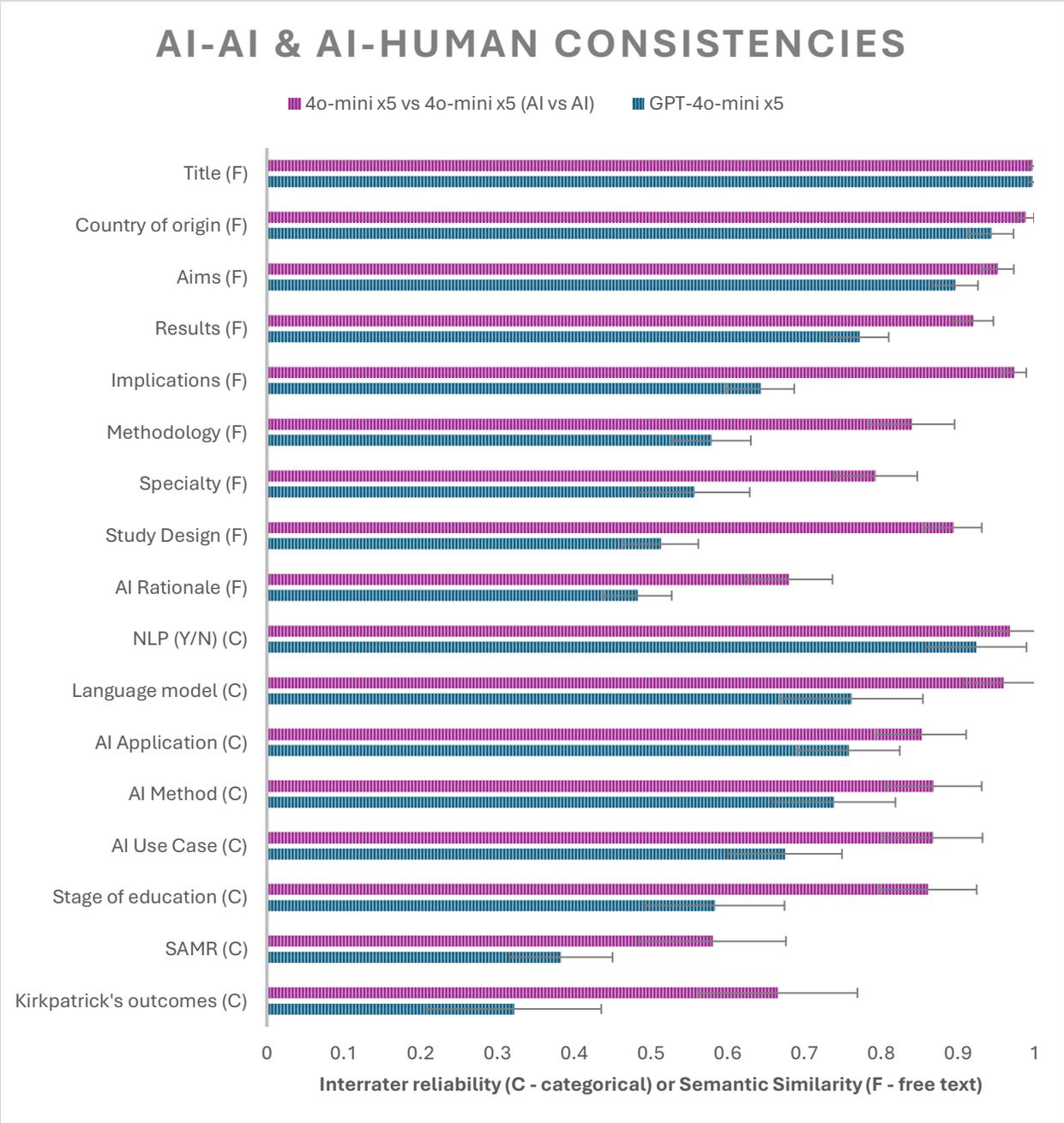

**Figure 4:** *AI-AI consistency also demonstrates a question-specific pattern*. The bar graph shows the consistency (IRR for categorical C, and semantic similarity for free text F) between two sets of AI extraction responses (4o-mini x5 vs 4o-mini x5) to the same question/publication pairings over the entire dataset shown in Figure 2. This is compared to AI-human consistency over the same dataset (4o-mini x5 vs Human). The brackets show the 95% confidence intervals for the consistency measures. The AI-AI consistency reflects the range of possible responses produced by an LLM to repeated re-generation of the same question/publication pairing. Similar to human-human and AI-human consistencies, the AI-AI consistency is lower for questions with greater expected interpretability (Kirkpatrick, SAMR, AI rationale), again demonstrating a question-specific pattern.

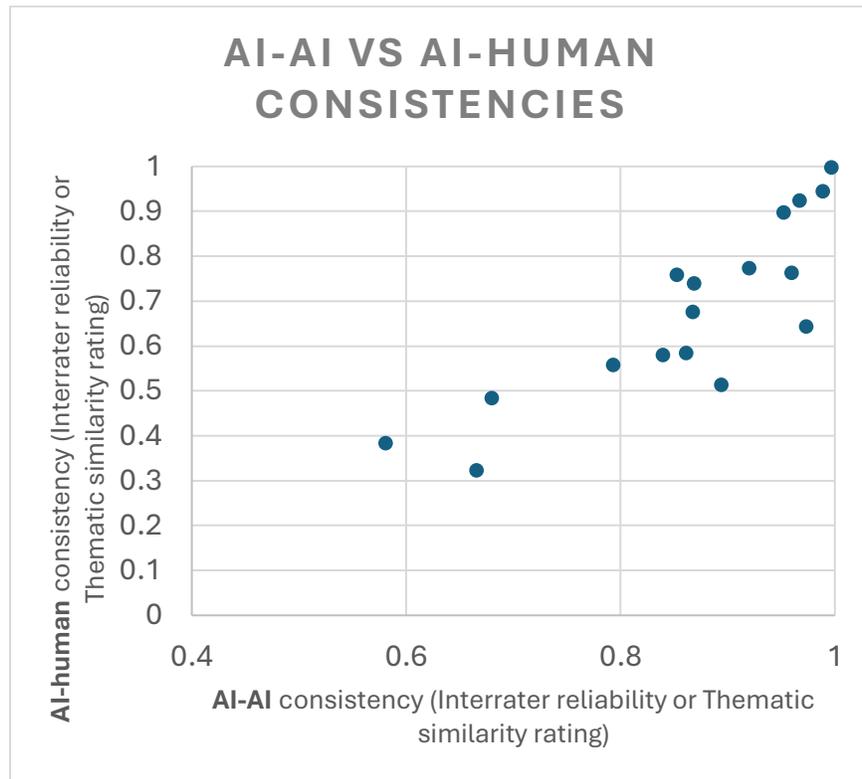

**Figure 5:** *AI-human consistency and AI-AI consistency are correlated* (Pearson correlation 0.85 ± 0.14), demonstrating a similar question-specific pattern. Each point represents the AI-human and AI-AI consistencies for a given extraction question, across all publications. This suggests that AI models may be able to identify the interpretability of extraction questions without human input.

*Identifying "hallucinations:" unpacking AI-human discordance*

Having identified inconsistencies between AI and human responses to extraction questions, we sought to uncover the sources of this discordance to assess for potential AI hallucinations (Table 4 and Figure 6). We found that discordance between AI and human responses could be broadly classified as errors of *inaccuracy* or *insufficiency* in extraction. We characterized inaccuracy as: *omission* (O) of entire element(s) being extracted, *major misclassification* (C) of the publication in the categorization schemas provided in the extraction prompts, and *factual lapses* (F) where the extractor asserted that the publication stated something that it did not. We characterized insufficiency as a less severe form of inaccuracy, involving: *specificity deficit* (S) relating to a lack of detail or partial completeness (not amounting to complete omission), and *minor misclassification* (M) (e.g. assignment to adjacent categories on an ordinal categorization scheme).

Applying this extraction error classification scheme to each discordant extraction case (i.e. question/publication pairs with low to moderate AI-human consistency), we found that AI extraction made 48 inaccuracy errors (out of 17 x 187 = 3179 total cases, for a rate of 1.51%) involving major omissions (O) or major misclassifications (C), but without any factual lapses (F). Human extractors made 139 inaccuracy errors (4.37%) involving major misclassifications (M) and factual lapses (F), but only one major omission (O). As such, compared to AI, humans appeared to be more likely to assign an inapplicable categorization rather than to omit a key category – i.e. ***human classification appeared to have higher sensitivity but lower specificity than AI classification***, and vice versa.

Errors of insufficiency were exhibited equally by AI and human extractors (135 vs 146 cases, respectively, or rates of 4.25% vs 4.59%). They appeared to primarily involve specificity deficits (S) and minor misclassifications (M), at similar rates for similar extraction questions when comparing AI to human extractors. Examples of specificity deficits (S) included: correctly identifying one of a study's analytic strategies while omitting others, and partial descriptions of methodology, aims, or results. Minor misclassifications (M) included assignment of Kirkpatrick or SAMR levels one level higher or lower than the most justified assignment. AI and human extractors appeared to perform roughly equally with respect to the rate of such errors.

Overall, 769 (24.2%) AI-human response pairs could be classified as discordant (Figure 6). Of these, 583 cases (18.3%) did not involve errors of accuracy (factual lapses (F) or major omissions (O)). 330 (10.3%) cases exhibited discordant AI and human responses that yet were complete and justifiable independent of each other. As such, this 10.3% rate of discordant-but-justifiable responses likely reflects extraction question interpretability (multiple answers to the same question). The remaining 253 (8.0%) discordant cases represented insufficient AI or human responses, reflecting a difference in level of detail or exhibiting minor misclassifications (Figure 6). Such minor misclassifications could also be considered reflecting question interpretability.

**Table 4:** Analysis of discordant AI-human responses.

| Question | AI Extraction | | Human Extraction | | Disc / Fact |
| --- | --- | --- | --- | --- | --- |
| | Inaccuracy | Insufficiency | Inaccuracy | Insufficiency | |
| Q_1 Title | C(1) Misinterpreted page header as title | S(1) Omitted part of the title (missing subtitle) | (0) | S(2) Used an abbreviated or paraphrased title, omitting key phrases | 4 / **1** |
| Q_3 Country of origin | (0) | S(6) Listed only some of the countries represented or oversimplified multiple to a single region | F(5) Added countries not present (e.g., listing "U.S." when the study was solely in the UK) or misinterpreted the author's affiliation as the study setting | S(6) Listed some but not all countries or mistakenly combined related regions | 16 / **2** |
| Q_5 Aims | (0) | S(5) Captured part of the aim but left out secondary objectives or qualifiers | F(1) Stated an aim not supported by the paper – claiming the study sought to develop AI tools when it only evaluated existing tools | S(9) Captured one component of the aim but missed another (e.g., focusing on teaching without assessment) | 29 / **15** |
| Q_7 Results | (0) | S(3) Summarized only part of the results – the primary outcome but none of the secondary findings | F(1) Drew unwarranted conclusions, stated numerical results that weren't in the manuscript | S(7) Mentioned one or two findings but omitted the rest or combined outcomes incorrectly | 83 / **73** |
| Q_8 Implications | (0) | S(2) Provided a general implication but left out specifics | F(7) Included implications (e.g., future application) not explicitly discussed in the study | S(3) Included only one aspect of the paper's implications | 75 / **63** |
| Q_18 Methodology | C(3) Misclassified the study's method or stated no methodology was given when it was | S(6) Gave a partial description mentioning data sources but omitting analysis techniques | O(1) Incorrectly stated that no methods used | S(6) Provided some correct methodology details but omitted key steps or misnamed one component | 36 / **20** |
| Q_4 Specialty | (0) | S(13) Labeled a broad area instead of the specific specialty or missed that multiple specialties were involved | SC(28) Assigned the wrong specialty (e.g., "surgery" instead of "radiology") or listed multiple specialties when the paper focused on one | S(7) Picked one correct specialty but missed additional ones, or said "multiple" when a primary specialty was specified | 60 / **14** |
| Q_6 Study Design | C(1) misclassified the design (e.g., survey vs. randomized controlled trial). | S(6) Provided an incomplete description – e.g., called it "observational" without specifying prospective vs retrospective | CF(5) Mislabeled the design type (e.g., describing a cross-sectional survey as a cohort study) or added design features not present | SM(22) Noted the correct general design but omitted qualifiers or miscategorized prospective vs retrospective | 76 / **42** |
| Q_16 AI Rationale | O(3) Stated that no rationale was provided when it was | S(7) Mentioned part of the rationale but omitted others | (0) | SM(1) Captured one correct rationale but missed another, or conflated the rationale with the study's aim | 41 / **30** |
| Q_12 NLP (Y/N) | C(2) Stated NLP was used when it wasn't, or vice versa | (0) | C(5) Stated NLP was used when it wasn't, or vice versa | M(2) Misinterpreted mentions AI models as NLP use, or vice versa | 9 / **0** |
| Q_14 Language model | C(1) Discussed study design instead of naming any language model | S(1) Listing one model (e.g., "ChatGPT") but missing others mentioned | CF(8) Named a model when none was used, or conflated the paper's methodology with a specific model | SM(3) Mentioned one correct model but omitted others; sometimes confused | 28 / **15** |

| | | | | language models with NLP in general | |
|---|---|---|---|---|---|
| Q_11 AI Application | OM(4) Assigned an incorrect application (e.g., "assessment" instead of "teaching") or saying "N/A" when one existed | S(3) Named only one application when the paper addressed multiple (e.g., focusing on simulation without noting assessment) | C(8) Assigned an incorrect application | S(5) Noted one application correctly but omitted another | 47 / **27** |
| Q_9 AI Method | OM(5) Named the wrong methodology (e.g., calling a Bayesian model a "deep neural network") or incorrectly stated that no AI method was used | S(12) Listed only some of the methods applied (e.g., machine learning but not deep learning) | CF(23) Assigned an incorrect category, often overly applying Deep Learning (DL), or adding a method not mentioned | SM(9) Listed one correct method but omitted others or conflated different categories | 47 / **3** |
| Q_13 AI Use Case | M(1) Misclassified a survey study about AI use as if the study itself applied AI for content generation when it did not | SM(8) Provided partially correct but imprecise labels, instead substituting broader, related, or tangential use cases | C(6) Misclassified AI as providing clinical, procedural, or assessment functions when the manuscripts explicitly described them as without those roles | M(10) Partial but imprecise matches to the actual reported use case, mislabeling AI's role when manuscripts described narrower or different analytic purposes | 43 / **19** |
| Q_2 Stage of education | OM(3) Labeled the wrong stage or incorrectly claimed no stage | S(14) Listed only one stage when the study involved multiple | C(16) Assigned a different stage or incorrectly assigned "mixed" | S(10) Captured one correct stage but left out the second, or misinterpreted dual-stage descriptions | 43 / **2** |
| Q_17 SAMR | OM(16) Selected an inappropriately high or low SAMR level without clear supporting evidence, or claimed no level | M(45) Chose a level adjacent to the most appropriate one | C(13) Misclassified the SAMR tier, choosing extremes without clear supporting evidence | M(34) Provided a near-adjacent level (e.g., one step above or below the most appropriate) or hedged by listing multiple levels | 97 / **1** |
| Q_19 Kirkpatrick's outcomes | OM(8) Assigned an incorrect evaluation level or stated "N/A" when a level was explicit | M(4) Chose a level adjacent to the most appropriate one or lacked certainty | C(11) Selected a level that would have required measuring outcomes not evaluated | M(10) Chose a level neighboring an appropriate level, or listed two adjacent levels | 35 / **3** |
| Total | 48 | 135 | 139 | 146 | 769 / **330** |
| Rate | 1.51% | 4.25% | 4.37% | 4.59% | 24.2% **18.3%** |

Disc = total number of discrepancies between AI and human responses for each question
Fact = total number of discrepant cases where both the AI and human responses were justifiable and factual, indicative of an *interpretable question*. The total number of interpretable questions could be considered as this value plus the number of minor misclassifications.
(#) = the number of responses characterized by the inaccuracy/insufficiency
Coding of inaccuracy/insufficiency:
- **O**mission = failure to identify a major element being extracted (more severe than below)
- Major mis**C**lassification = true misclassification, improper equating
- **F**actual lapse = assertion that the publication contains facts that it does not actually state
- **S**pecificity deficit = insufficient detail, lack of completeness, oversimplification, or generalization
- **M**inor misclassification = classification into a level neighboring (or similar to) the most appropriate level

Note that the number of inaccurate cases per row did not necessarily add up to the number of AI-human discrepancies, since in some cases there were both AI and human inaccuracies (when the sum across the row exceeded the total number of discrepancies) and some cases where AI and human were both accurate (when the sum across the row was less than the total number of discrepancies).

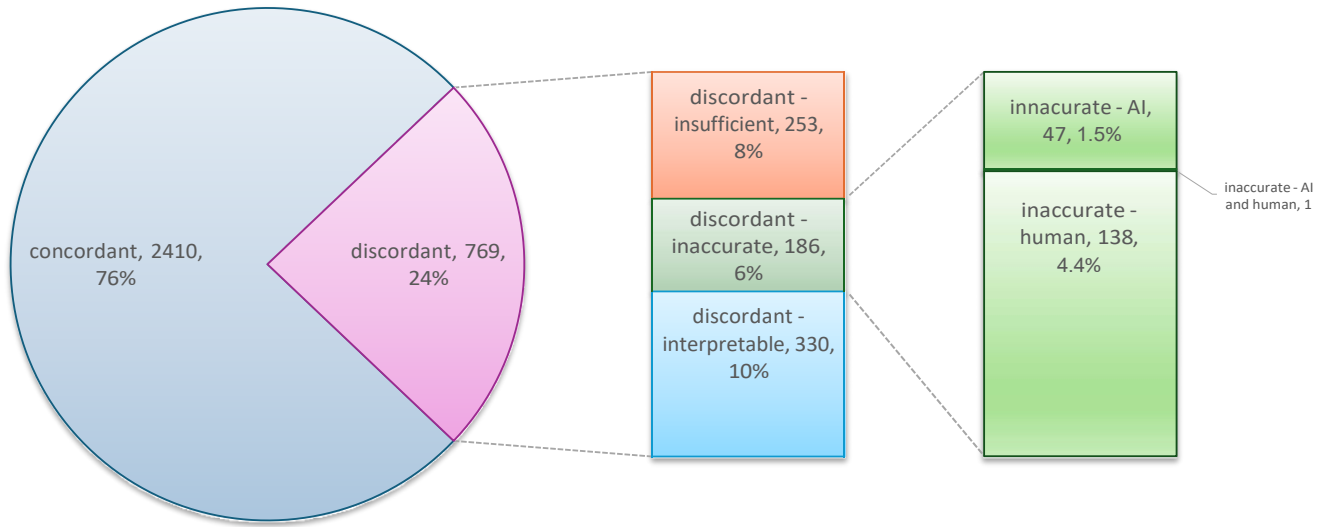

**Figure 6:** Discordance analysis (above) and error rates (below). ABOVE: 769 (24%) of 3179 total question/publication pairs demonstrated discordance between AI and human responses to extraction questions. If these, 330 (10.3%) represented AI and human responses that were both factually accurate relative to the publication text, reflecting question interpretability. 253 (8%) represented cases where AI and/or human responses were insufficient, reflecting a mismatch in detail. 186 (6%) represented cases where AI and/or human responses were inaccurate, of which 47 (1.5%) represented AI inaccuracies and 138 (4.4%) human inaccuracies (with one case sharing both AI and human inaccuracy). BELOW: AI extraction had an error rate of 1.5% for inaccurate responses (omissions and misclassifications) and 4.3% for insufficient responses (lack of specificity/detail or minor misclassifications). Notably AI did not demonstrate factual lapses. Human extraction had an error rate of 4.4% for inaccurate responses (misclassifications and factual lapses) and 4.6% for insufficient responses. AI errors were characterized by omissions without factual lapses, while the opposite was true for human errors, suggesting a difference in prioritization of sensitivity and specificity.

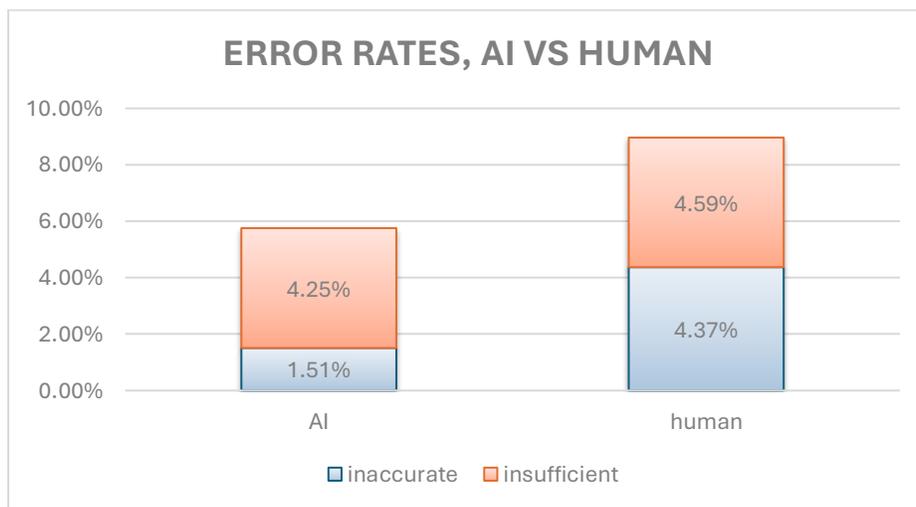

## Discussion

We assessed AI extraction accuracy by comparing AI responses to human responses, and by comparing them both to their source publications. We found that extraction questions with high AI-human consistency were typically concrete and required minimal interpretation (e.g. title, country of origin, aims and results) (Table 3). By contrast, lower consistency arose when questions required more interpretation (Kirkpatrick's levels, SAMR) or were concrete but not described in explicit terms (e.g. study design). Quantitatively, we measured AI-human consistency as interrater reliability for categorical items and thematic similarity for free-text items, which showed a stable, question-specific pattern independent of the AI model (Figure 2). Notably, for interpretation-heavy questions, human-human consistency was no higher than AI's ability to reproduce human answers (Figure 3). Taken together, question-specific consistency recurred across AI-human, human-human, and AI-AI comparisons.

*Interpretation vs hallucination*. We distinguished hallucination (asserting content absent from the source) from interpretation (reasonable variation) when responding to an extraction question. If hallucination had dominated, we would have expected broadly uniform hallucination rates across questions and manuscripts (Asgari et al., 2025). Hallucinations would have affected AI-human and AI-AI (Figures 4 & 5) comparisons more than human-human comparisons, because trained human extractors should rarely invent content absent from the text. By contrast, if interpretation dominated, we would expect strong question- or publication-specific dependence and similar patterns across AI-human, human-human, and AI-AI comparisons. Empirically we observed the latter, suggesting that interpretation – not hallucination – primarily explains variability in responses.

We performed an analysis of AI and human errors (Figure 6) to triangulate this finding. Analysis of discordant AI and human responses (769/3179 = 24.2%) revealed that AI inaccuracies were responsible for only a small number of cases (1.51%), while interpretive differences – multiple correct answers, unequal levels of detail, and minor classification disagreements – were responsible for the bulk of AI-human discord (18.3%).

*Error profiles and operating thresholds*. Error analysis separated inaccuracies (wrong or unjustified claims, including omissions and major misclassifications) from insufficiencies (thematically correct but lacking detail or granularity, including minor misclassifications). AI errors were more often omissive without factual lapses, whereas human errors more often involved misclassifications and factual assertions not supported by the text. Humans were nearly three times more likely to commit inaccuracy errors than AI (4.37% vs 1.51%), and we observed no cases of AI asserting facts absent from the source. This pattern suggests different implicit operating points on a sensitivity-specificity trade-off: AI favored specificity (withhold when uncertain), while humans favored sensitivity (commit when a classification seems plausible). Framed as an implicit receiver operating characteristic (ROC) curve, the two systems appear to use different decision thresholds (Pintea & Moldovan, 2009; Youngstrom, 2014). Calibrating an AI collaborator may thus involve threshold tuning to align with human coders (and vice versa) rather than combating an intrinsic tendency to "hallucinate."

*Implications for AI trustworthiness.* An extraction question's interpretability is central to deciding whether to accept an AI response. Highly interpretable questions are also those most susceptible to researcher positionality and reflexivity, which an AI – with opaque positionality – cannot easily declare. Moreover, interpretability is not always obvious at question design time and may go unrecognized when relying solely on AI (which typically does not flag that multiple defensible answers exist) or solely on a homogeneous human team (with limited reflexive diversity).

*Platform contribution and workflow.* To make interpretability visible, MAKMAO includes multi-pass extraction and consistency diagnostics. For a given manuscript, the platform (i) repeats each question multiple times and (ii) compares responses with each other and with justifying source text. Operationally, low AI-AI consistency functions as an early warning flag: it signals either (a) high interpretability (multiple defensible readings) or (b) possible hallucination/inaccuracy when answers cannot be justified from the source. MAKMAO makes this distinction explicit by pairing variability with source-linked justifications (Figure 1b) and an error-justification view (Figure S2).

Because AI-AI consistency correlates with AI-human consistency (Figure 4), repeated AI runs provide a low-cost proxy for human benchmarking. We propose deploying this capability in KS during extraction-question development to prototype and stress-test items (Figure 7). Importantly, variability does not equate to interpretability by default: variability may also reflect underspecified or ambiguous question wording. A practical refinement loop is: (1) run repeated extractions; (2) revise the prompt with definitions/anchors/exemplars; (3) re-run; and (4) compare dispersion. If variability shrinks after clarification, the issue was ambiguity; if variability persists, the question is likely to be genuinely interpretable and may warrant explicit handling (e.g., multiple codes, positionality notes). Iterating with AI thus helps sharpen questions before full human (and perhaps, AI) extraction.

**Figure 7**. *A suggested workflow for developing an extraction question* leveraging repeated AI extractions. How to distinguish interpretability from undesired ambiguity and hallucinations.

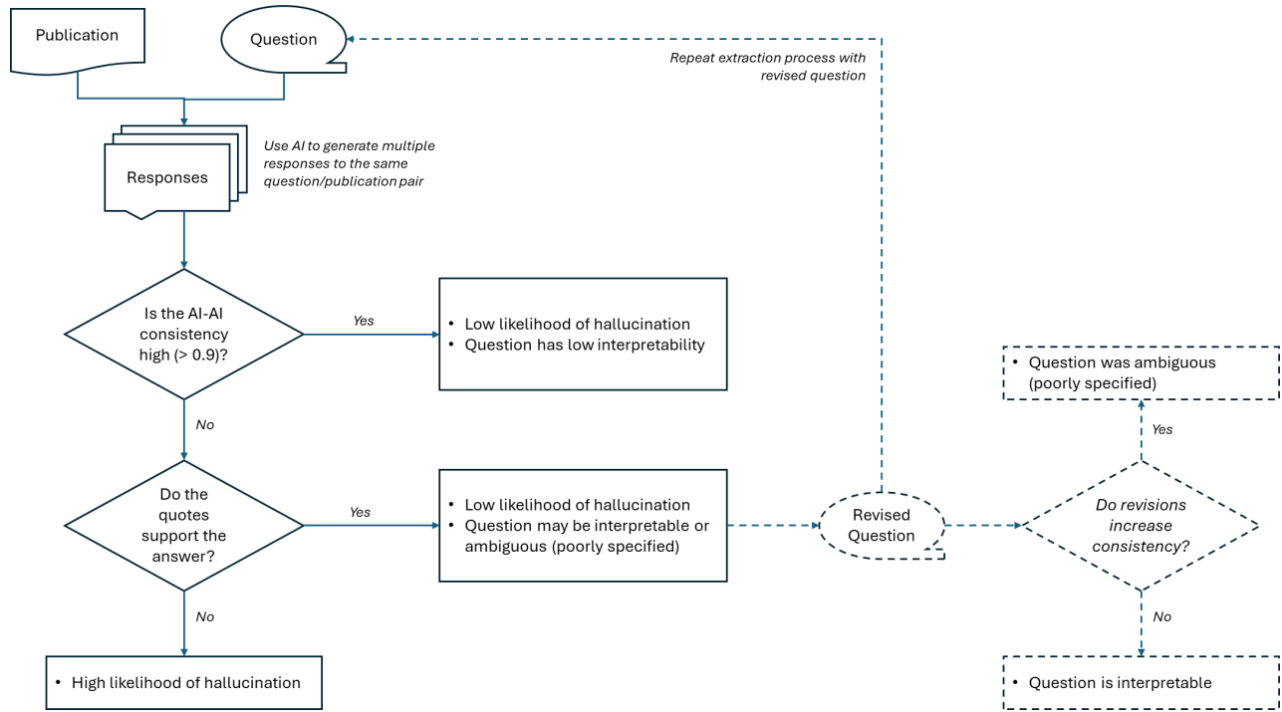

## Limitations and further directions

Our dataset focused on AI in HPE, which may limit generalizability. Future work should (a) examine how explicit and implicit biases shape answers to interpretive questions for both AI and humans; (b) characterize AI positionality, including how training data and prompting steer which viewpoints are expressed; and (c) further differentiate interpretability from nonspecific ambiguity due to poor question specification – for example, by contrasting distinct, defensible theme clusters (interpretability) with diffuse, unstable themes (ambiguity). Finally, our error-checking could be integrated as an internal QC gateway in iterative answer generation to improve accuracy.

Beyond publications, similar methods could support narrative assessment scoring/feedback (replacing "publications" with learner texts and "extraction questions" with rubric prompts) and clinical documentation extraction, with validity assessed using the same comparative design.

## Conclusion

If HPE's scholarly discourse functions as a form of distributed cognition, then knowledge synthesis is its reflective consolidation. How much should we lean on AI to facilitate this reflection? Trustworthiness is one ingredient in the decision to use AI; so too are researchers' motivations and desired degree of engagement in the extraction process. Over-reliance can yield bland, low-insight summaries that miss the "productive struggle" by which conceptual linkages emerge (Bowen, 2009; Cook et al., 2025). Under-reliance risks forgoing opportunities to systematize extraction and improve efficiency in a process that averages 67.3 weeks to complete (Borah et al., 2017) and ~881 person-hours, 24% of which are spent on data extraction (Pham et al., 2018). Extraction questions encode a KS's theories, hypotheses, and analytic lenses. Designing them is labor-intensive and often resource-constrained. By using repeated AI extractions to assess extraction question interpretability, researchers can target questions that need clearer definitions and decide where human interpretive work is essential. Exploring question interpretability will not only help researchers to reduce ambiguities, but also to recognize where interpretability may be both inevitable, desired, and worthy of further investigation. These advantages move AI beyond a mere time-saver towards a true collaborative partner that enhances researchers' ability to develop insights from data.

## Disclosures

All authors are inventors on a U.S. provisional patent application for the MAKMAO extraction workflow described in this manuscript. The authors have no other conflicts of interest to disclose.

## References

Ammanath, B. (2022). *Trustworthy AI: A Business Guide for Navigating Trust and Ethics in AI*. Wiley.

Asgari, E., Montaña-Brown, N., Dubois, M., Khalil, S., Balloch, J., Yeung, J. A., & Pimenta, D. (2025). A framework to assess clinical safety and hallucination rates of LLMs for medical text summarisation. *Npj Digital Medicine*, *8*(1), 274. https://doi.org/10.1038/s41746-025-01670-7

Baker-Brunnbauer, J. (2023). *Trustworthy Artificial Intelligence Implementation*. Springer International Publishing. https://doi.org/10.1007/978-3-031-18275-4

Bernard, N., Sagawa, Y., Bier, N., Lihoreau, T., Pazart, L., & Tannou, T. (2025). Using artificial intelligence for systematic review: the example of elicit. *BMC Medical Research Methodology*, *25*(1). https://doi.org/10.1186/s12874-025-02528-y

Bolaños, F., Salatino, A., Osborne, F., & Motta, E. (2024). Artificial intelligence for literature reviews: opportunities and challenges. *Artificial Intelligence Review*, *57*(10), 259. https://doi.org/10.1007/s10462-024-10902-3

Borah, R., Brown, A. W., Capers, P. L., & Kaiser, K. A. (2017). Analysis of the time and workers needed to conduct systematic reviews of medical interventions using data from the PROSPERO registry. *BMJ Open*, *7*(2), e012545. https://doi.org/10.1136/bmjopen-2016-012545

Bowen, G. A. (2009). Document analysis as a qualitative research method. *Qualitative Research Journal*, *9*(2), 27–40. https://doi.org/10.3316/QRJ0902027

Cook, D. A., Ginsburg, S., Sawatsky, A. P., Kuper, A., & D'Angelo, J. D. (2025). Artificial Intelligence to Support Qualitative Data Analysis: Promises, Approaches, Pitfalls. *Academic Medicine*. https://doi.org/10.1097/ACM.0000000000006134

Gin, B. C., Holzhausen, Y., Khursigara-Slattery, N., Chen, H. C., Schumacher, D. J., & ten Cate, O. (2024). Theoretical foundations of trust and entrustment in health professions education. In O. ten Cate, V. Burch, H. Chen, F. Chou, & M. Hennus (Eds.), *Entrustable Professional Activities and Entrustment Decision-Making in Health Professions Education* (pp. 35–50). Ubiquity Press. https://doi.org/10.5334/bdc.d

Gin, B. C., O'Sullivan, P. S., Hauer, K. E., Abdulnour, R.-E., Mackenzie, M., ten Cate, O., & Boscardin, C. K. (2025). Entrustment and EPAs for Artificial Intelligence (AI): A Framework to Safeguard the Use of AI in Health Professions Education. *Academic Medicine*, *100*(3), 264–272. https://doi.org/10.1097/ACM.0000000000005930

Gordon, M., Daniel, M., Ajiboye, A., Uraiby, H., Xu, N. Y., Bartlett, R., Hanson, J., Haas, M., Spadafore, M., Grafton-Clarke, C., Gasiea, R. Y., Michie, C., Corral, J., Kwan, B., Dolmans, D., & Thammasitboon, S. (2024). A scoping review of artificial intelligence in medical education: BEME Guide No. 84. *Medical Teacher*, *46*(4), 446–470. https://doi.org/10.1080/0142159X.2024.2314198

Grimmelikhuijsen, S. (2023). Explaining Why the Computer Says No: Algorithmic Transparency Affects the Perceived Trustworthiness of Automated Decision-Making. *Public Administration Review*, *83*(2), 241–262. https://doi.org/10.1111/puar.13483

Hsieh, H.-F., & Shannon, S. E. (2005). Three approaches to qualitative content analysis. *Qualitative Health Research*, *15*(9), 1277–1288. https://doi.org/10.1177/1049732305276687

Maggio, L. A., Costello, J. A., Norton, C., Driessen, E. W., & Artino Jr, A. R. (2020). Knowledge syntheses in medical education: a bibliometric analysis. *Perspectives on Medical Education*, *10*(2), 79–87. https://doi.org/10.1007/S40037-020-00626-9

Mayer, R. C., & Mayer, B. M. (2024). *A research agenda for trust* (R. C. Mayer & B. M. Mayer, Eds.). Edward Elgar Publishing.

Park, Y. S., Hyderi, A., Bordage, G., Xing, K., & Yudkowsky, R. (2016). Inter-rater reliability and generalizability of patient note scores using a scoring rubric based on the USMLE Step-2 CS format. *Advances in Health Sciences Education*, *21*(4), 761–773. https://doi.org/10.1007/s10459-015-9664-3

Pham, B., Bagheri, E., Rios, P., Pourmasoumi, A., Robson, R. C., Hwee, J., Isaranuwatchai, W., Darvesh, N., Page, M. J., & Tricco, A. C. (2018). Improving the conduct of systematic reviews: a process mining perspective. *Journal of Clinical Epidemiology*, *103*, 101–111. https://doi.org/10.1016/j.jclinepi.2018.06.011

Pintea, S., & Moldovan, R. (2009). The Receiver-Operating Characteristic (ROC) analysis: Fundamentals and applications in clinical psychology. *Journal of Cognitive and Behavioral Psychotherapies*, *9*(1), 49–66.

Spillias, S., Ollerhead, K., Andreotta, M., Annand-Jones, R., Boschetti, F., Duggan, J., Karcher, D., Paris, C., Shellock, R., & Trebilco, R. (2024). *Evaluating Generative AI to Extract Qualitative Data from Peer-Reviewed Documents* (pp. 1–23). https://doi.org/10.21203/rs.3.rs-4922498/v1

U.S. Department of Health & Human Services, H. (2021). HHS Trustworthy Artificial Intelligence (AI) Playbook. In *U.S. Department of Health & Human Services* (Issue September). https://www.hhs.gov/sites/default/files/hhs-trustworthy-ai-playbook.pdf

Xu, F., Hao, Q., Zong, Z., Wang, J., Zhang, Y., Wang, J., Lan, X., Gong, J., Ouyang, T., Meng, F., Shao, C., Yan, Y., Yang, Q., Song, Y., Ren, S., Hu, X., Li, Y., Feng, J., Gao, C., & Li, Y. (2025). *Towards Large Reasoning Models: A Survey of Reinforced Reasoning with Large Language Models*. https://doi.org/10.48550/arXiv.2501.09686

Youngstrom, E. A. (2014). A primer on receiver operating characteristic analysis and diagnostic efficiency statistics for pediatric psychology: We are ready to ROC. *Journal of Pediatric Psychology*, *39*(2), 204–221. https://doi.org/10.1093/jpepsy/jst062

# Appendix 1: Extraction Prompts

```
FIRST GEN PROMPTS:
```

What was the first author of the paper's last name. Give this name only as your brief answer, without other text.

What year was the paper published. Give the year only in your short answer.

What is the title of the paper.

what is the name of the journal the paper was published in? Give the journal name only.

What is the type of publication? Provide your answer as one of the following only: perspective, innovation, or article.

What was the stage of education? Provide your answer only as one or more of the following abbreviations: UME (undergraduate medical education), GME (graduate medical education), CPD (continuing professional development)

In what country(ies) was the study performed?

What was the medical specialty (if applicable). Provide your answer only as the name of the specialty(ies) only, i.e. radiology, surgery, pediatrics, internal medicine, etc.

What are the aims, goals and objectives of the paper? Provide you answer as one sentence, starting with "To determine," "To assess," "To describe," or a similar "To …" phrase.

What is the study design of the paper? (provide your answer as 6 or fever words, for example: survey, case report, two group comparison, pre/post knowledge & attitudes, model development, cross sectional study, comparison of LLM performance, comparative study, etc.)

Summarize the results of the study in at least 2 complete sentences.

What did the study propose as implications for future educational practice, policy or research? Answer in at least two complete sentences.

Did the study apply AI to a specific task in medical education or practice (Yes), or did the study only investigate knowledge/attitudes about AI without applying it (No)? (Answer Yes or No only)

If the study did not apply AI to a specific use, answer "N/A", otherwise answer: What did the study apply AI to, or what was the study's focus of use for AI? Provider your answer only as one or more of the following: admissions/selection, teaching/instruction, assessment, clinical reasoning, automation of case/procedure logs, performance of LLMs on a medical exam.

If the study did not apply AI to a specific use, answer "N/A", otherwise answer: What AI method was used? Provide your answer only as one or more of the following: traditional ML (machine learning), DL (deep learning).

If the study did not apply AI to a specific use, answer "N/A", otherwise answer: Did the study utilize natural language processing (NLP)? Answer yes or no only.

If the study did not apply AI to a specific use, answer "N/A", otherwise answer: What was the study's AI use case? Provider your answer only as one or more of the following: predictive model, sentiment analysis, bias evaluation, chatbot, clinical guidance, personalized learning platforms, content generation, data labelling, VPS (virtual patient simulator), performance analytics, performance guidance, data analytics, summative assessment completion.

If the study did not apply AI to a specific use, answer "N/A", otherwise answer: If a paper used an AI language model or LLM please provide the name(s) of such models.

If the study did not apply AI to a specific use, answer "N/A", otherwise answer: What was the study's rationale for using AI? Answer in one sentence only.

If the study did not apply AI to a specific use, answer "N/A", otherwise answer: In regards to the task that AI used to perform in the study, does the AI's performance represent: a Substitution (or replication), an Augmentation (or enhancement), a Modification (redesign to do the same thing), or a Redefinition (creation of a new task) of that original task?

If the study did not apply AI to a specific use, answer "N/A", otherwise answer: How did the study implement AI? (Answer in 1 sentence only)

If the study did not apply AI to a specific use, answer "N/A", otherwise answer: If the study measured a learning outcome, what Kirkpatrick level did the measured outcomes represent? (Use the numerical numbers from the Kirkpatrick's Four Levels of Training Evaluation)

FINAL (REFINED) PROMPTS:

Q_1: What is the title of the paper.

Q_2: What was the stage of education? Provide your answer only as one or more of the following abbreviations: UME (undergraduate medical education – i.e. medical school, nursing school, etc), GME (graduate medical education – i.e. residency, internship, or fellowship), CPD (continuing professional development – i.e. professional practice)

Q_3: In what country(ies) was the study performed?

Q_4: What medical specialty does the paper apply to (if applicable)? Provide your answer only as the name of the specialty(ies) only, i.e. radiology, surgery, pediatrics, internal medicine, etc.

Q_5: What are the aims, goals and objectives of the paper? Provide you answer as one sentence, starting with "To determine," "To assess," "To describe," or a similar "To …" phrase.

Q_6: Identify the study design and categorize it into a concise label based on the methodology used. The output should be a single phrase that best describes the study type, aligning with standard research classifications. The response should be similar in format to these example categories: Survey, Retrospective cohort study, Randomized Control Trial (RCT), Cross-sectional study, Model development & validation, Innovation and outcome evaluation, Qualitative study (Focus Groups, Interviews), Two-group comparison, Observational cohort study, Pre-post intervention assessment, Delphi study, etc.

Q_7: Summarize the results of the study in at least 2 complete sentences.

Q_8: State what the authors proposed as the study's implications for future educational practice, policy or research. Answer in at least two complete sentences.

Q_9: What AI method was used? Provide your answer only as one or more of the following: traditional ML (machine learning), DL (deep learning), or AI as a concept (if AI is not applied but is the subject of a course, or a study about attitudes towards AI).

Q_10: Did the study apply AI to a specific task in medical education or practice (Yes), or did the study only investigate knowledge/attitudes about AI without applying it (No)? (Answer Yes or No only)

Q_11: If the study did not apply AI to a specific use, answer "N/A", otherwise answer: What area of medical education did the study apply AI to? Provider your answer only as one or more of the following: admissions/selection, teaching/instruction, assessment, clinical reasoning, automation of case/procedure logs, performance of LLMs on a medical exam.

Q_12: If the study did not apply AI to a specific use, answer "N/A", otherwise answer: Did the study utilize natural language processing (NLP)? Answer yes or no only.

Q_13: If the study did not apply AI to a specific use, answer "N/A", otherwise answer: What analytic or technological purpose did the study use AI to perform? Provider your answer only as one or more of the following: predictive model, sentiment analysis, bias evaluation, chatbot, clinical guidance, personalized learning platforms, content generation, data labelling, VPS (virtual patient simulator), performance analytics, performance guidance, data analytics, summative assessment completion.

Q_14: If the study did not apply AI to a specific use, answer "N/A", otherwise answer: If a paper used an AI language model or LLM please provide the name(s) of such models.

Q_15: If the study did not apply AI to a specific use, answer "N/A", otherwise answer: What software and/or programming languages did the study use?

Q_16: If the study did not apply AI to a specific use, answer "N/A", otherwise answer: What was the study's rationale for using AI? In other words, what was their justification for using AI to accomplish a task in medical education? Answer in one sentence only.

Q_17: If the study did not apply AI to a specific use, answer "N/A", otherwise answer: Use the SAMR framework to describe how the study applied AI to the medical education task. Answer only one of the following: Substitution (using AI as a direct substitute for traditional methods without functional change), Augmentation (using AI as a substitute with functional improvements), Modification (using AI enables significant task redesign), or Redefinition (using AI allows for the creation of an entirely new, previously inconceivable task)

Q_18: If the study did not apply AI to a specific use, answer "N/A", otherwise answer: What methodology or approach was used in this study, including how AI was applied? (Answer in 1-2 sentences)

Q_19: If the study did not apply AI to a specific use, or did not explicitly measures learning outcomes from from human participants in an educational intervention or intervention, answer "N/A", otherwise answer: Determine which Kirkpatrick evaluation level the humans' outcomes are most consistent with (do not report Kirkpatrick levels for AI performance, only human performance). Provide only the number of the level per the following key: 1 - Reaction: the paper focuses on participants' immediate responses, satisfaction, or engagement (e.g., survey feedback), 2 - Learning: the paper measures increases in knowledge, skills, or attitudes (e.g., pre/post-tests, quizzes), 3 - Behavior: the paper evaluates how participants apply new skills or knowledge in practice (e.g., observations, self-assessments), 4 - Results: the paper reports broader organizational impacts or outcomes (e.g., performance metrics, ROI).

# Appendix 2:

Instructions for accessing the platform: TBA

# Appendix 3: Sensitization of GPT-5 to the Gordon and Daniels's Scoping Review

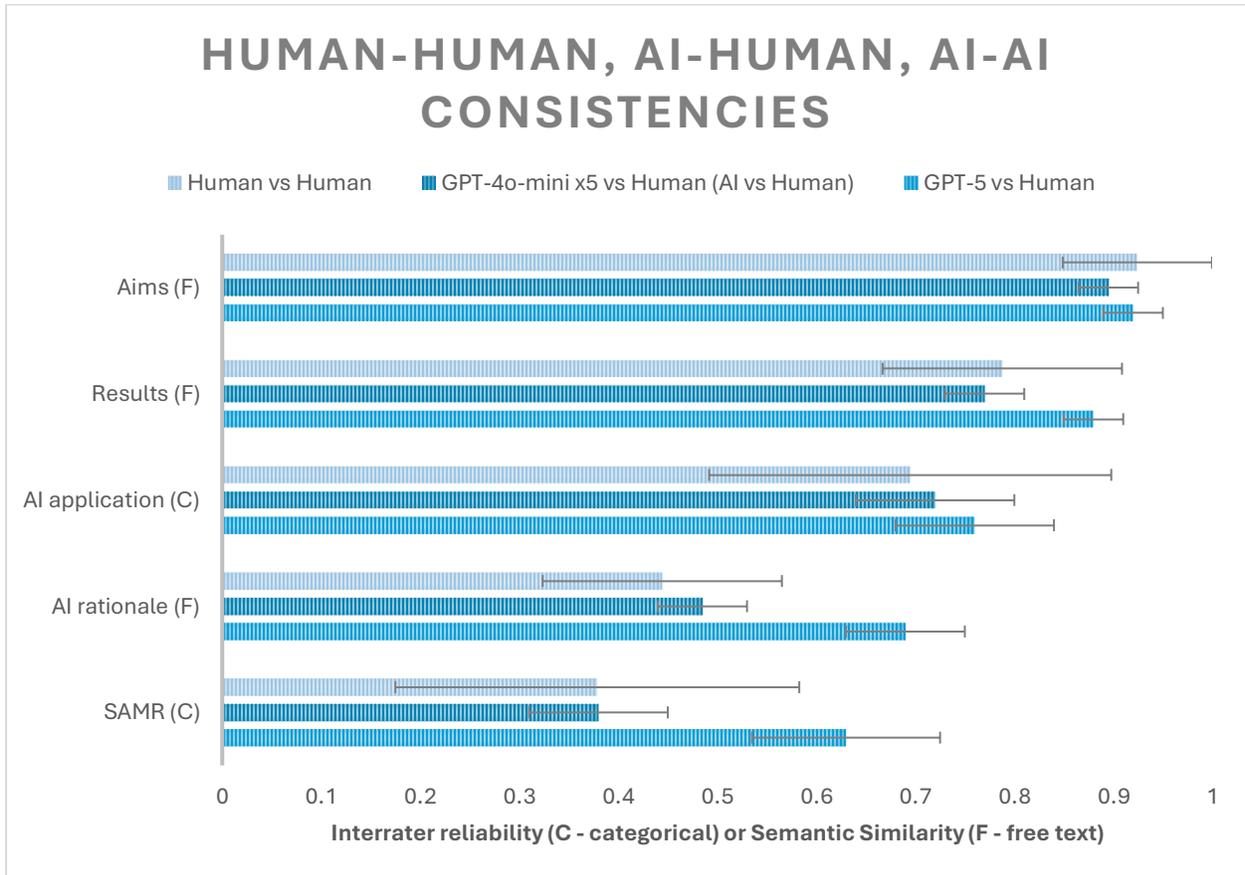

The AI-human consistency of the LRM (GPT-5) was significantly (p<0.05) higher than the human-human consistency for the AI rationale and SAMR questions, while still demonstrating high correlation to the human-human pattern (Pearson correlation = 0.97). While it is possible that GPT-5 answered questions in a way that was consistently more aligned with the original KS's human extractors (than ourselves as human extractors), a more likely explanation is that the previously published extraction table (and associated KS publication) was used in the training of GPT-5, given that its training cutoff date (6/2024) was after that KS was published. Indeed, when we asked GPT-5 if it knew of this KS study, it returned: "Yes — from my training data, I know of the **Morris Gordon & Michelle Daniel rapid scoping review** on AI in medical education. It's the **BEME Guide No. 84**, published in *Medical Teacher* in early 2024. The review systematically mapped the literature on AI applications across undergraduate, postgraduate, and continuing medical education, following Arksey & O'Malley's scoping review framework." As such, we excluded GPT-5's extraction performance from further analysis. This issue does not appear to be the case for the other LLMs we studied here that were either trained before this KS was published or did not endorse knowledge of it.